\documentclass[aps,pra,twocolumn,superscriptaddress,showpacs]{revtex4-1}
\usepackage{amsmath}
\usepackage{amsfonts}
\usepackage{amssymb}
\usepackage{graphicx}
\usepackage{color}
\usepackage{subfigure}

\usepackage{hyperref}

\newcommand{\ket}[1]{\left|#1\right\rangle}
\newcommand{\bra}[1]{\left\langle #1\right|}
\newcommand{\abssq}[1]{\left|#1\right|^2}
\newcommand{\expect}[1]{\left\langle #1 \right\rangle}

\begin{document}

%\preprint{}

\title{Collapse and Revival and Cat States with an $N$ Spin System}


\author{Shane Dooley}
\email[]{pysd@leeds.ac.uk}
%\thanks{}
\affiliation{Quantum Information Science, School of Physics and Astronomy, University of Leeds, Leeds LS2 9JT, United Kingdom.}

\author{Francis McCrossan}
\affiliation{Quantum Information Science, School of Physics and Astronomy, University of Leeds, Leeds LS2 9JT, United Kingdom.}

\author{Derek Harland}
\affiliation{Department of Mathematical Sciences, Loughborough University, Loughborough, Leics LE11 3TU, United Kingdom.}

\author{Mark J. Everitt}
\affiliation{Department of Physics, Loughborough University, Loughborough, Leics LE11 3TU, United Kingdom.}

\author{Timothy P. Spiller}
\affiliation{Quantum Information Science, School of Physics and Astronomy, University of Leeds, Leeds LS2 9JT, United Kingdom.}


\date{\today}

\begin{abstract}
We discuss collapse and revival of Rabi oscillations in a system comprising a qubit and a ``big spin'' (made of $N$ qubits, or spin-$1/2$ particles). We demonstrate a regime of behaviour analogous to conventional collapse and revival for a qubit-field system, employing spin coherent states for the initial state of the big spin. These dynamics can be used to create a cat state of the big spin. Even for relatively small values of $N$, states with significant potential for quantum metrology applications can result, giving sensitivity approaching the Heisenberg limit. 
\end{abstract}

\pacs{03.65.Ta, 03.67.-a, 05.50.+q, 42.50.Xa }

%\keywords{}

\maketitle

\section{Introduction}

Collapse and revival of the Rabi oscillations of a qubit, or two-level atom, coupled to a field mode \cite{Jay-63, Ebe-80} provide a remarkable and much-discussed illustration of the quantum nature of the composite qubit-field system. The collapse of the Rabi oscillations arises through qubit-field entanglement, yet half way to the revival of the oscillations the field and qubit disentangle again, with the quantum information in their initial states effectively swapped into their counterpart system \cite{Gea-90}. This enables, for example, the generation of superpositions of two (approximate) coherent states -- sometimes called a cat state -- of the field \cite{Gea-91}. The addition of further qubits has led to the phenomena of entanglement sudden death \cite{Yu-04, Yu-09, Jar-09}, collapse and revival of entanglement \cite{Jar-09} and cat-swapping \cite{Jar-10}.

In this work we discuss collapse and revival of the Rabi oscillations of a qubit coupled to a ``big spin'' made of $N$ qubits, or spin-$1/2$ particles. We demonstrate that preparation of the $N$ spins in a spin coherent state---the analogue of a coherent state for a field mode---facilitates collapse and revival phenomena. We isolate a regime of parameter space where conventional looking collapse and revival occurs and thus a superposition of two (approximate) spin coherent states of the big spin -- a spin cat state -- emerges prior to revival, although we note that different and interesting forms of revival dynamics occur in other parameter regimes.

Quantum metrology \cite{Lee-02,Gio-04, Gio-11} is an emerging quantum information application where entangled quantum resources are employed to measure an unknown external potential or field that through interaction generates a phase in the state of the quantum resources. Enhanced measurement precision---beyond the ``Standard Quantum Limit'' achieved using the resources one-by-one, or classically---is possible. In principle, precision all the way down to the ``Heisenberg limit'' can be attained. Cat states are one form of non-classical resource with significant potential for metrology, with spin cats enabling enhanced magnetic field sensing \cite{Xio-10}. We therefore assess the ability of the states produced through spin collapse and revival for quantum metrology, and demonstrate that accuracy close to that of the Heisenberg limit could be attained.

\section{Collapse and Revival and Cat States}
\subsection{The Jaynes-Cummings Model}
\begin{figure}
\includegraphics[width=80mm, scale=0.62]{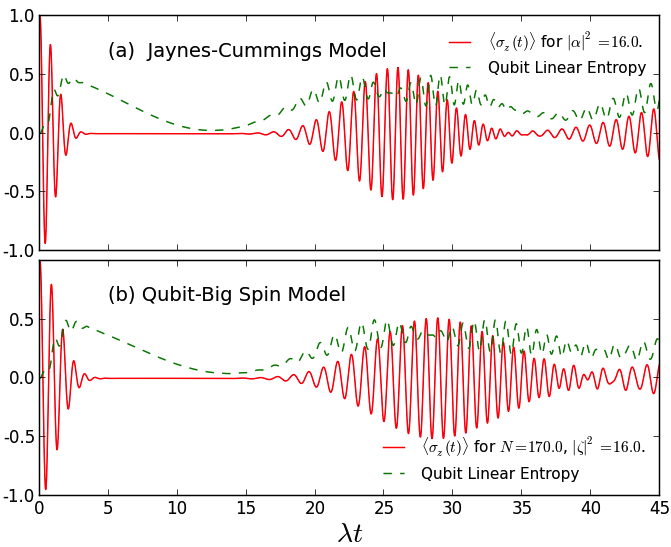}
\caption{(colour online). When $\frac{\abssq{\zeta}}{N} \ll 1 \ll N$, there is a correspondence between the qubit-big spin model and the JC model. Here we see collapse and revival of $\expect{\sigma_z (t)}$ (solid red line) in (a) the JC model, and (b) the qubit-big spin model. \label{fig:JCvsBS}}
\end{figure}

We begin our discussion with the familiar qubit-field dynamics. The Jaynes-Cummings (JC) model \cite{Jay-63} for the interaction of a bosonic field mode with a qubit, or two level atom, is described by the Hamiltonian\begin{equation} \hat{H}_{\footnotesize{\mbox{JC}}} = \omega \hat{a}^{\dagger}\hat{a} + \frac{\Omega}{2}\hat{\sigma}_z + \lambda \left( \hat{a}^{\dagger}\hat{\sigma}_- + \hat{a} \hat{\sigma}_+ \right), \label{eq:JCH}\end{equation} where $\hat{a}^{\dagger}$ and $\hat{a}$ are the field mode creation and annihilation operators respectively, and $\hat{\sigma}_z = \ket{0}\bra{0}-\ket{1}\bra{1}$, $\hat{\sigma}_{+} = \ket{0}\bra{1}$ and $\hat{\sigma}_{-} = \ket{1}\bra{0}$ are qubit operators. Assuming resonance ($\Omega = \omega$), if the qubit is initially in the state $\ket{0}$ and the field initially in a coherent state $\ket{\alpha} = \ket{|\alpha |e^{i\phi}}$ we see the well-known collapse and revival of oscillations of $\expect{\hat{\sigma}_z (t)}$ [solid red line, figure \ref{fig:JCvsBS} (a)] and the dip in the qubit entropy at half of the revival time [dashed green line, figure \ref{fig:JCvsBS} (a)]. 

Gea-Banacloche \cite{Gea-90,Gea-91} has made an insightful analysis of these features. For example, he has shown that when $\abssq{\alpha}\gg 1$, there are two initially orthogonal qubit states, $\ket{D_{\pm} (0)}=\frac{1}{\sqrt{2}}\left( \ket{0} \pm e^{-i\phi}\ket{1} \right)$, that -- to a good approximation -- evolve without entangling with the field mode: \begin{equation} \ket{D_{\pm}(0)}\ket{\alpha} \to  \ket{D_{\pm} (t)} e^{\mp i\lambda t \sqrt{\hat{a} \hat{a}^{\dagger}} }  \ket{\alpha} . \label{eq:D0} \end{equation} He has also shown that at a particular time \begin{equation} t_0 = \frac{\pi}{\lambda}\sqrt{ \bra{\alpha}\hat{a}^{\dagger}\hat{a} \ket{\alpha} } = \frac{\pi |\alpha|}{\lambda}, \label{eq:fieldt0} \end{equation} which is equal to half the revival time, these states coincide: $\ket{D_{+}(t_0)} = \ket{D_{-}(t_0)}$. This is known as the \emph{attractor state} of the qubit and it is independent of the initial qubit state. Since any pure state of the qubit can be written as a superposition of $\ket{D_{+}(0)}$ and $\ket{D_{-}(0)}$, it follows that any initial qubit state will converge to the attractor state at $t_0$. This explains the dip in qubit entropy at half the revival time in figure \ref{fig:JCvsBS} (a). Since the state of the composite qubit-field system is pure at all times, the field must also be in a pure state at $t_0$. For a judicious choice of initial qubit state, the field is in a cat state at this time \cite{Gea-91,Buz-92}.

\subsection{The Qubit-Big Spin Model}

Instead of the interaction of a field mode with a two level atom, we consider the interaction of a system of $N$ spin-$\frac{1}{2}$ particles, or qubits, (which we refer to as ``the big spin'') with a single spin-$\frac{1}{2}$ particle (a qubit) via the Hamiltonian \begin{equation}  \hat{H} = \omega \left( \hat{J}_z + \frac{N}{2} \right) + \frac{\Omega}{2} \hat{\sigma}_z  + \frac{\lambda}{\sqrt{N}}\left(  \hat{J}_{+}\hat{\sigma}_{-} + \hat{J}_{-}\hat{\sigma}_{+}  \right), \label{eq:spinH} \end{equation} where $\hat{J}_z \equiv \frac{1}{2} \sum_{i=1}^{N} \hat{\sigma}_{z}^{(i)}$ and $\hat{J}_{\pm} \equiv \sum_{i=1}^{N} \hat{\sigma}_{\pm}^{(i)}$ are operators that act on the big spin and $\hat{\sigma}_{z}^{(i)} = \ket{\uparrow^{(i)}}\bra{\uparrow^{(i)}} -  \ket{\downarrow^{(i)}}\bra{\downarrow^{(i)}}$ acts on the individual spins that make up the big spin. The constant term $\frac{\omega N}{2}$ in (\ref{eq:spinH}) is not really necessary, but is convenient because it shifts the spectrum of the big spin Hamiltonian $\hat{J}_z$ so that its ground state eigenvalue is zero.

We also introduce the operator $\hat{J}^2 = \hat{J}_{x}^2 + \hat{J}_{y}^2+ \hat{J}_{z}^2$ [which commutes with our Hamiltonian (\ref{eq:spinH})]. \emph{Dicke states} $\ket{j,n-j}$ are simultaneous eigenstates of $\hat{J}_z$ and $\hat{J}^2$ with eigenvalues $n-j$ and $j(j+1)$ respectively. In what follows we restrict to the $j=\frac{N}{2}$ eigenspace of the $N$ spin system. This is an $N+1$ dimensional subspace for which the Dicke states $\ket{\frac{N}{2}, n-\frac{N}{2}}$ ($n\in\left\{0,1,...,N\right\}$) form a basis.

A \emph{spin coherent state} \cite{Are-72, Rad-71} in the $j=\frac{N}{2}$ eigenspace is a state in which each of the $N$ spins is in the same pure state. Parameterised by the complex number $\zeta$, this spin coherent state is: \small\begin{equation}  \ket{N,\zeta} = \bigotimes_{i = 1}^N \left(  \frac{1}{\sqrt{1 + \abssq{\zeta}}} \ket{\downarrow^{(i)}} + \frac{\zeta}{\sqrt{1 + \abssq{\zeta}}} \ket{\uparrow^{(i)}}  \right).  \label{eq:scs} \end{equation}\normalsize

There is an equivalent representation of this spin coherent state in terms of Dicke states: $\ket{N,\zeta} = \sum_{n=0}^N C_n \ket{ \frac{N}{2}, n - \frac{N}{2} } $ where 
\begin{equation} C_{n} =  \frac{1}{\left(  1 + \abssq{\zeta}   \right)^{N/2}} \sqrt{ \frac{N!}{(N-n)!n!} }\: \zeta^n  . \label{eq:spincoh}\end{equation}

We consider the qubit-big spin system evolving by Hamiltonian (\ref{eq:spinH}) where the big spin is initially in the spin coherent state $\ket{N, \frac{\zeta}{\sqrt{N}}}$ where $\zeta$ has been scaled by a factor of $1/\sqrt{N}$. (This scaling turns out to be useful when we consider $N\to\infty$.)

%We say that the big spin is initially in a \emph{spin coherent state} $\ket{N,\zeta}$ \cite{Are-72, Rad-71}. An example of a spin coherent state of the $N$ spin system is one in which each of the $N$ spins is in the same pure state. This state is parameterised by the complex number $\zeta$: \small\begin{equation}  \ket{N,\zeta} = \bigotimes_{i = 1}^N \left(  \frac{1}{\sqrt{1 + \abssq{\zeta}/N}} \ket{\downarrow^{(i)}} + \frac{\zeta /\sqrt{N}}{\sqrt{1 + \abssq{\zeta}/N}} \ket{\uparrow^{(i)}}  \right).  \label{eq:scs} \end{equation}\normalsize There is an equivalent representation of this spin coherent state in terms of the angular momentum states, $\ket{\frac{N}{2}, n - \frac{N}{2}} \equiv \ket{n}$, which are simultaneous eigenstates of $\hat{J}_z$ (with eigenvalue $n- N/2$) and the total angular momentum operator $\hat{J}^2 = \hat{J}_{x}^2 + \hat{J}_{y}^2+ \hat{J}_{z}^2$ [with eigenvalue $\frac{N}{2}(\frac{N}{2}+1)$]. In terms of these states, the spin coherent state is $\ket{N,\zeta} = \sum_{n=0}^N C_n \ket{ n } $ where \begin{equation} C_{n} =  \frac{1}{\left(  1 + \frac{\abssq{\zeta}}{N}   \right)^{N/2}} \sqrt{ \frac{N!}{(N-n)!n!} } \left( \frac{\zeta}{\sqrt{N}} \right)^n  . \label{eq:spincoh}\end{equation}

%  \begin{equation} \small \ket{\frac{N}{2}, n - \frac{N}{2}} \equiv \binom{N}{n}^{-1/2} \displaystyle\sum_{\substack{\mbox{ \footnotesize permu-}\\ \mbox{tations }} } \ket{\downarrow^{\otimes ( N-n ) }\uparrow^{\otimes n}} , \end{equation}

Plotted in figure \ref{fig:JCvsBS} (b) (again assuming resonance $\Omega = \omega$, qubit initially in $\ket{0}$, and initial spin coherent state $\ket{N, \frac{\zeta}{\sqrt{N}}}$ with $\zeta =4$, $N =170$) are $\expect{\hat{\sigma}_z (t)}$ and the linear entropy of the qubit. The oscillations in $\expect{\hat{\sigma}_z}$ undergo a collapse and revival that is very similar to the collapse and revival in the JC model. 

%\subsubsection{Approximation in the $1\ll \abssq{\zeta}\ll N$ domain}

The similarities between our qubit-big spin system and the JC model can be understood by looking at the $N\to\infty$ limit of the $N$ spin system. To see the connection, we consider an embedding of the $N$ spin system in the Hilbert space of the field mode by the linear map $f$ that takes the Dicke state $\ket{\frac{N}{2},n-\frac{N}{2}}$ to the Fock state $\ket{n}$:

\begin{eqnarray}  & & \quad f\ket{\frac{N}{2},n-\frac{N}{2}} = \ket{n}  ; \\  & & f\ket{\frac{N}{2},n-\frac{N}{2}}\bra{\frac{N}{2},m-\frac{N}{2}}f^{\dagger} = \ket{n}\bra{m} . \end{eqnarray} %With this mapping the spin state $\ket{\frac{N}{2}, -\frac{N}{2}} = \ket{\downarrow\downarrow...\downarrow}$ corresponds to the `vacuum state

Restricting to the $j=\frac{N}{2}$ eigenspace of the $N$ spin system and taking the $N\to\infty$ limit we find (shown in appendix \ref{app:A}) that

\begin{eqnarray}\lim_{N\to\infty} f \frac{\hat{J}_-}{\sqrt{N}} f^{\dagger} = \hat{a} \quad ; \quad \lim_{N\to\infty} f \frac{\hat{J}_+}{\sqrt{N}} f^{\dagger} = \hat{a}^{\dagger} ; \label{eq:HPA} \\  \lim_{N\to\infty} f \left( \hat{J}_z + \frac{N}{2} \right) f^{\dagger} = \hat{a}^{\dagger}\hat{a} . \qquad \label{eq:HPB} \end{eqnarray}

% Show these in appendix??

% Alternatively, one can view the relations above as the $N\to\infty$ limit of the Holtein-Primakoff transformations.

 Combining these equations we see that our Hamiltonian (\ref{eq:spinH}) is the same as the Jaynes-Cummings Hamiltonian (\ref{eq:JCH}) in the $N\to\infty$ limit: $ \lim_{N\to\infty} f \hat{H} f^{\dagger} = \hat{H}_{JC} $.
 
 Moreover, one can use the Poisson Limit Theorem (see appendix {\ref{app:A}}) to show that, in the $N\rightarrow\infty$ limit our initial spin coherent state $\ket{N,\frac{\zeta}{\sqrt{N}}}$ is mapped onto the field mode coherent state:
\begin{equation}  \lim_{N\to\infty}f \ket{N,\frac{\zeta}{\sqrt{N}}} = e^{-\abssq{\zeta}/2}\sum_{n=0}^{\infty} \frac{\zeta^n}{\sqrt{n!}}\ket{n}  . \label{eq:Poisson} \end{equation}

In the $N\to\infty$ limit we see collapse and revival of $\expect{\hat{\sigma}_z}$ because both our big spin initial state and Hamiltonian are mathematically the same as those that result in collapse and revival in the field-qubit interaction. 

%This correspondence between our big spin-qubit model and the JC model is exact in the $N\to\infty$ limit. If $N$ is finite, it is not exact but is very good when $\frac{\abssq{\zeta} }{N} \ll 1\ll N$. In other words, if many spin half particles are initially aligned with the same polarisation (such that $\frac{\abssq{\zeta} }{N} \ll 1\ll N$) and are allowed to interact with a single qubit via Hamiltonian (\ref{eq:spinH}), then this system evolves -- to a good approximation -- like the JC model. If $\abssq{\zeta}\approx N$ the correspondence between the qubit-big spin model and the JC model breaks down.
 
This correspondence between our big spin-qubit model and the JC model is exact in the $N\to\infty$ limit. If $N$ is finite, it is not exact. In particular, when $N$ is finite the bosonic commutation relation $\left[ \hat{a}, \hat{a}^{\dagger} \right] = \hat{\mathbb{I}}$ is not satisfied by the corresponding big spin operators: \begin{equation} \left[ \frac{\hat{J}_{-}}{\sqrt{N}} , \frac{\hat{J}_{+}}{\sqrt{N}} \right] = \hat{\mathbb{I}} - \frac{2}{N}\left( \hat{J}_{z} + \frac{N}{2} \right) \end{equation} If the correspondence with the JC model is to hold approximately, we have -- as a minimum requirement (since in the bosonic case $\expect{ \left[ \hat{a}, \hat{a}^{\dagger} \right] } = 1$ for any state) -- that $\expect{\left[ \frac{\hat{J}_{-}}{\sqrt{N}} , \frac{\hat{J}_{+}}{\sqrt{N}} \right] } \approx 1$ for the initial spin coherent state of the big spin. Since $\bra{N,\zeta} \left( \hat{J}_{z} + \frac{N}{2} \right) \ket{N,\zeta} =  \frac{\abssq{\zeta}}{1+\abssq{\zeta}/N}$, this leads to the requirement that $\abssq{\zeta}\ll N$. In other words, if many spin half particles are initially aligned with the same polarisation (such that $\abssq{\zeta} \ll N$) and are allowed to interact with a single qubit via Hamiltonian (\ref{eq:spinH}), then this system evolves -- to a good approximation -- like the JC model. If $\abssq{\zeta}\approx N$ the correspondence between the qubit-big spin model and the JC model breaks down.

We make use of the correspondence when $\abssq{\zeta}\ll N$ to propose a method of creating spin cat states. Since Gea-Banacloche's approximation for the field mode is valid when $ 1 \ll \abssq{\alpha}$, we expect the same approximation to be valid for our qubit-big spin system when $1 \ll \abssq{\zeta} \ll N$. In that parameter regime we say that initial qubit states $\ket{D_{\pm} (0)}=\frac{1}{\sqrt{2}}\left( \ket{0} \pm e^{-i\phi}\ket{1} \right)$ evolve without entangling with the big spin: \begin{equation} \ket{D_{\pm} (0)} \ket{N, \zeta} \to  \ket{D_{\pm} (t)} e^{\mp i\lambda t \sqrt{\hat{J}_- \hat{J}_+ /N} }  \ket{N, \zeta} , \label{eq:noent}\end{equation} and that at time \begin{equation} t_0 = \frac{\pi}{\lambda} \sqrt{ \bra{N,\frac{\zeta}{\sqrt{N}}} \left( J_z + \frac{N}{2} \right) \ket{N,\frac{\zeta}{\sqrt{N}}} } = \frac{\pi |\zeta |}{\lambda\sqrt{1 + \frac{\abssq{\zeta}}{N}}} \label{eq:spint0} \end{equation} the qubit is in an attractor state. [Equations (\ref{eq:noent}) and (\ref{eq:spint0}) correspond to equations (\ref{eq:D0}) and (\ref{eq:fieldt0}) for the field mode, but with $\hat{a}$, $\hat{a}^{\dagger}$, $\hat{a}^{\dagger}\hat{a}$ and the initial coherent state replaced by the corresponding big spin operators and the initial spin coherent state via equations (\ref{eq:HPA}, \ref{eq:HPB}, \ref{eq:Poisson}).] The big spin is, at this time, in a pure state that will depend on the initial state of the qubit. For qubit initially $\ket{0} = \frac{1}{\sqrt{2}} \left(  \ket{D_+ (0)} + \ket{D_- (0)}  \right)$, the big spin is, at time $t_0$, in the state \small\begin{equation} \ket{\psi_{\zeta} } = \frac{1}{\sqrt{2\mathcal{M}}} \Big[  e^{-i\lambda t_0 \sqrt{ \frac{ \hat{J}_-\hat{J}_+ }{N} } }\ket{N,\zeta}  + e^{+i\lambda t_0 \sqrt{\frac{\hat{J}_-\hat{J}_+ }{N} } }\ket{N,\zeta}   \Big] , \label{eq:cat}\end{equation}\normalsize  a spin cat. The $\mathcal{M}$ has been introduced to maintain normalisation of $\ket{\psi_{\zeta}}$ since $e^{-i\lambda t_0 \sqrt{\hat{J}_- \hat{J}_+ /N} }\ket{N,\zeta}$ and $e^{+i\lambda t_0 \sqrt{\hat{J}_- \hat{J}_+ /N} }\ket{N,\zeta}$ are, in general, not orthogonal to each other.

\begin{figure}
\centering
	\includegraphics[width=80mm]{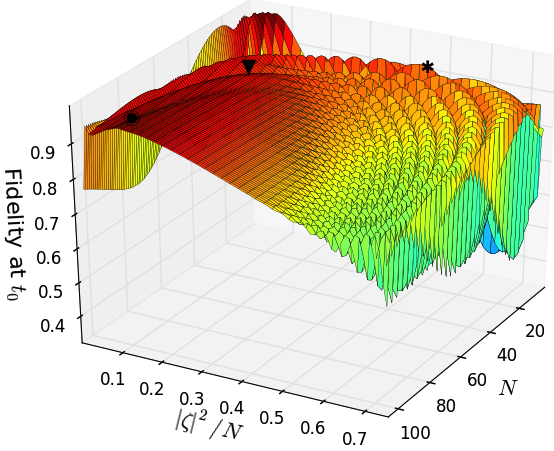} %{FidelityN=4to100_8.png} 
	\caption{(colour online). Red indicates areas of high fidelity. Fidelity is high when $1 \ll \abssq{\zeta} \ll N$, but also around $\abssq{\zeta}/N \approx 0.5$ for certain small values of $N$.}
	\label{fig:3d}
\end{figure}

Figure \ref{fig:3d} shows $F = \sqrt{ \bra{\psi_{\zeta}} \rho_{\mbox{\footnotesize BS}} (t_0) \ket{\psi_{\zeta}}  }$, the fidelity of $\ket{\psi_{\zeta}}$ against $\rho_{\mbox{\footnotesize BS}}(t_0)$, the (exact) reduced big spin state at $t_0$, plotted against $\abssq{\zeta}/N$ for various values of $N$. As expected (given the correspondence between our big spin model and the JC model) the fidelity is high when $1\ll \abssq{\zeta}\ll N$. At $N=100$ and $\abssq{\zeta}=6$, for example (marked by a black dot in figure \ref{fig:3d}), the fidelity at $t_0$ is high ($\sim 0.96$). This is telling us that the big spin system is close to the cat state $ \ket{\psi_{\zeta}} $ at $t_0$. 

\begin{figure}[]%[ht]
\centering
\subfigure{
    \includegraphics[width=35mm]{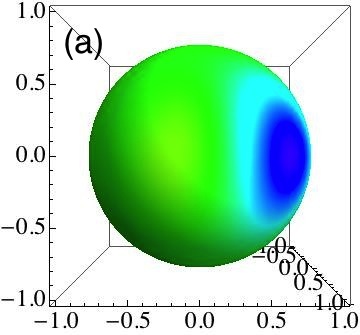}
   % \label{fig:SpWig5}
}
\subfigure{
	\includegraphics[width=35mm]{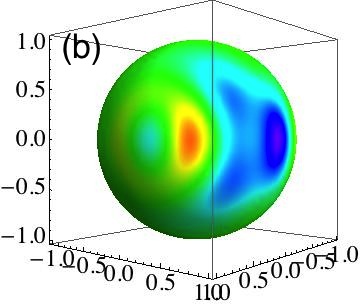}%{placeholder.jpg}
    %\label{fig:SpWig10}
}
\subfigure{
	\includegraphics[width=35mm]{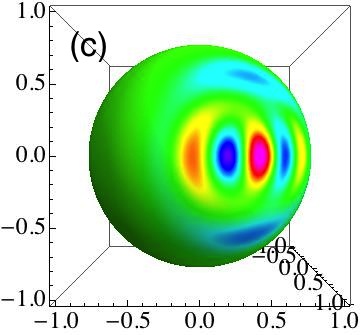}
%	\label{fig:SpWig20}
}
\subfigure{
	\includegraphics[width=35mm]{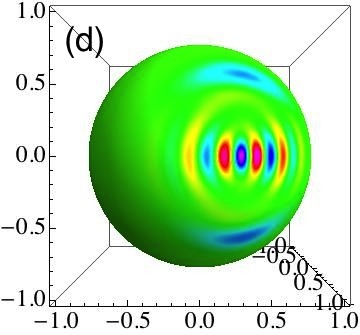}
%	\label{fig:SpWig40}
}
\caption[]{(colour online). Spin Wigner functions of $\rho_{\mbox{\footnotesize BS}}(t_0)$, the exact reduced big spin state at $t_0$. (a) $N=5$; (b) $N=12$, $\abssq{\zeta}=6$; (c) $N=20$, $\frac{\abssq{\zeta}}{N} = 0.16$; (d) $N=40$, $\frac{\abssq{\zeta}}{N} = 0.16$. Interactive figures showing the states in (b) and (c) are available as supplementary material \cite{Sup1}.}
\label{fig:SpWig}
\end{figure}

Interestingly, figure \ref{fig:3d} shows that this domain of high fidelity includes relatively small values of $N$. At $N=40$, for example (marked by a black triangle in figure \ref{fig:3d}), $F \sim 0.93$ at $t_0$ for $\abssq{\zeta} = 6$. To see that this is indeed a cat state, we plot in figure \ref{fig:SpWig}(d) its spin Wigner function \cite{Har-12}. A spin coherent state is represented by a circular blob in a spin Wigner plot. A superposition of spin coherent states would be represented by two circular blobs with interference fringes between them. Here, in figure \ref{fig:SpWig}(d), instead of circles, we have two crescent shapes with interference fringes between them -- clearly a cat state, although not quite a superposition of spin coherent states.

\begin{figure}[ht]%[ht]
\centering
    \includegraphics[width=80mm]{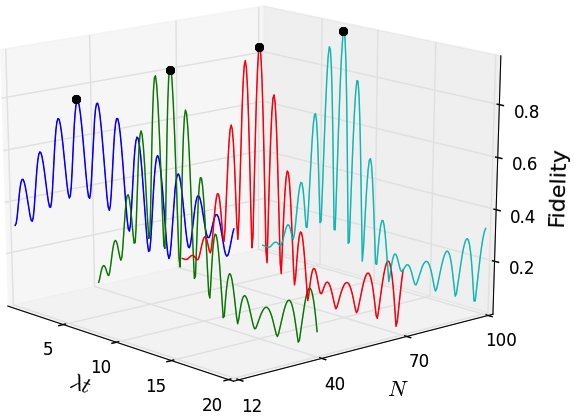}
    \caption{(colour online). Fidelity at $t_0$ is marked by a black dot. The fidelity around $t_0$ is highly oscillatory. ($\abssq{\zeta} = 6$ in each case.)}
    \label{fig:Fid t,N}
\end{figure}

Also of interest in figure \ref{fig:3d} are the ripples in the fidelity outside of our $1 \ll \abssq{\zeta}\ll N$ parameter regime, for example, for low $N$ around $\abssq{\zeta}/N \approx 0.5$. A cross section of figure \ref{fig:3d} at $\abssq{\zeta}/N = 0.5$ is plotted in figure \ref{fig:xsection} [the blue (upper) line]. These ripples are highly peaked for certain small values of $N$. At $N=12$, for example (marked by a black asterisk in figure \ref{fig:3d}), the fidelity to the cat state $\ket{\psi_{\zeta}} $ is $\sim 0.91$ at $t_0$. Figure \ref{fig:SpWig}(b) shows the spin Wigner function of this state. % Peaked, it seems, at intervals of 4 for $N$. Haven't checked yet but I bet that there are orthogonal cat states $\pm, i\pm$ that are close to the big spin state at this time for the other vales of $N$.

% The ripples show sensitivity to $N$, but not to $\zeta$. This is good because fixing $N$ shouldn't be a problem...

%Figure \ref{fig:3d} shows how the fidelity at $t_0$ varies with $\abssq{\zeta}$ and $N$. 

% For $N=100$, there is a broad range of values of $\abssq{\zeta}\ll N$ for which the fidelity at $t_0$ is high and slowly varying. This tells us that the fidelity at $t_0$ is not too sensitive to our choice of $\abssq{\zeta}$ for the initial spin coherent state. The fast oscillation of the fidelity in figure \ref{fig:Fid t,N}, however, tells us that the fidelity is sensitive to the interaction time. 

Figure \ref{fig:Fid t,N} shows $F = \sqrt{ \bra{\psi_{\zeta}} \rho_{\mbox{\footnotesize BS}} (t) \ket{\psi_{\zeta}}  }$, the fidelity of $\ket{\psi_{\zeta}}$ against $\rho_{\mbox{\footnotesize BS}}(t)$, the (exact) reduced big spin state, plotted against time for $N = 12, 40, 70, 100$ (with $\abssq{\zeta}=6$). Fidelity at $t_0$, marked by a black dot, is high in each case. As explained above, however, although the $N=12$ fidelity is high, it is in a different domain of high fidelity than $N=40,70,100$. 
% Have circle, triangle, square representing the three points on the fidelity landscape...

It is clear from figure \ref{fig:Fid t,N} -- since the fidelity is a highly oscillatory around $t_0$ -- that this method of generating a cat state is sensitive to the interaction time. Figure \ref{fig:3d}, on the other hand, shows that fidelity is not very sensitive to the initial spin coherent state parameter $\abssq{\zeta}$ when $1 \ll \abssq{\zeta} \ll N$.

The high fidelities that can be obtained for small values of $N$ are of interest for possible practical implementations of this cat state generation technique with existing technologies. For example, it has been shown that superconducting qubits can be coupled by a $\lambda \left( \sigma_+ \sigma_- + \sigma_- \sigma_+ \right)$ interaction Hamiltonian \cite{McD-05, Nis-07, Nee-10}. Our interaction Hamiltonian (\ref{eq:spinH}) is composed of $N$ such equal interactions with a central qubit. Alternatively, a superconducting phase qu$d$it \cite{Nee-09} (which emulates our $N+1$ dimensional $j=\frac{N}{2}$ subspace) might be coupled to a single superconducting qubit.

Another candidate system for realising a set of qubits separately coupled to a single system, without direct coupling to each other, is to use a superconducting resonator coupled (at the antinode of its microwave field) to a number of superconducting qubits, such as charge or transmon qubits. Experimental demonstrations have already been made with three or four superconducting qubits coupled to a resonator \cite{DiC-10,Ree-12}. In order to prevent multiple excitation of the resonator, thus limiting it to a qubit with just two effective levels, use of a non-linearity to detune other level separations might be appropriate. It is interesting to note that coherence times in these experimental superconducting systems are already at the point where collapse and revival phenomena can be observed due to a single-photon Kerr effect \cite{Kir-13}, giving real promise for the future application of these systems in metrology scenarios. 

%Alternatively, it has also been shown that superconducting qubits can be coupled by a $\sigma_+^{(i)} \sigma_- + \sigma_-^{(i)} \sigma_+$ interaction Hamiltonian \cite{McD-05,Nis-07}. Our interaction Hamiltonian (\ref{eq:spinH}) is composed of $N$ such equal interactions with a central qubit. %A superconducting phase quit, which emulates a quantum spin \cite{Nee-09}, might be coupled to a single superconducting qubit.

%Alternatively, two superconducting qubits can be coupled by the $\sigma_+^{(i)} \sigma_- + \sigma_-^{(i)} \sigma_+$ interaction Hamiltonian \cite{McD-05,Nis-07} that is necessary to build our Hamiltonian (\ref{eq:spinH}) for the interaction of $N$ qubits with a central qubit. 

%However, this method is limited by...unwanted direct interactions between the $N$ qubits as $N$ increases?

Aside from superconducting systems, another candidate system is any highly symmetric molecule that consist of $N$ spins equally coupled to a central spin. The trimethyl phosphite molecule, for example, has nine $^{1}H$ spins, all equally coupled to a single $^{31}P$ spin \cite{Jon-09}. The tetramethylsilane molecule has twelve $^{1}H$ spins equally coupled to a single $^{21}S$ spin \cite{Sim-10}. Using NMR techniques, entangled states of both of these molecules have already been generated for use as magnetic field sensors \cite{Jon-09, Sim-10}. 

In the next section we provide some analysis to quantify the usefulness of our spin cat states $\ket{\psi_{\zeta}}$ for magnetic field sensing.

% Maybe point out that what we call a cat state is not the same

% Regarding the sensitivity to the initial input parameters. Not very sensitive to the initial $\zeta$, sensitive to $N$ in the ripple domain, but this shouldn't be a problem, and sensitive to interaction time, but we assume resolution up to Rabi time

\section{Magnetic Field Sensing with Spin Cat States}

%We now try to quantify the usefulness of such a state for metrology or, in particular, for magnetic field sensing. 
A system of $N$ spin-$\frac{1}{2}$ particles initially in the cat state $\ket{\psi_{\zeta} }$ is allowed to interact with an unknown (classical) magnetic field, $\vec{B}=B_y$, via the Hamiltonian $\hat{H}=\gamma B_y \hat{J}_y  $, where $\gamma$ is the gyromagnetic ratio of our $N$ spin system. After a time $t$, our spin system is in the state $\ket{\psi_{\zeta} (\theta)}=e^{i\gamma B_y t \hat{J}_y}\ket{\psi_{\zeta}}= e^{i\theta \hat{J}_y}\ket{\psi_{\zeta} }$, where we have defined $\theta = \gamma t B_y$. Since $\gamma$ and $t$ are assumed to be known, estimating $\theta$ is the same as estimating $B_y$.  The precision $\Delta\theta$ with which we can estimate the parameter $\theta$ is bounded by the Cramer-Rao inequality \cite{Bra-94}, $(\Delta\theta)^2 \geq 1/\mathcal{F}$, where $\mathcal{F}$ is the quantum Fisher information. For ease of comparison between different values of $N$ we quantify precision by $N(\Delta\theta)^2 \geq N/\mathcal{F}$. Given that our $N$ spin system evolves unitarily and is initially in pure state $\ket{\psi_{\zeta}}$ we can write the quantum Fisher information as \begin{equation} \mathcal{F} = 4 \left(\Delta \hat{J}_y \right)^2 = \bra{\psi_{\zeta} } \hat{J}_{y}^2 \ket{\psi_{\zeta} } - \bra{\psi_{\zeta} }\hat{J}_{y}\ket{\psi_{\zeta} }^2 . \end{equation}

\begin{figure}
\includegraphics[width=75mm, scale=0.62]{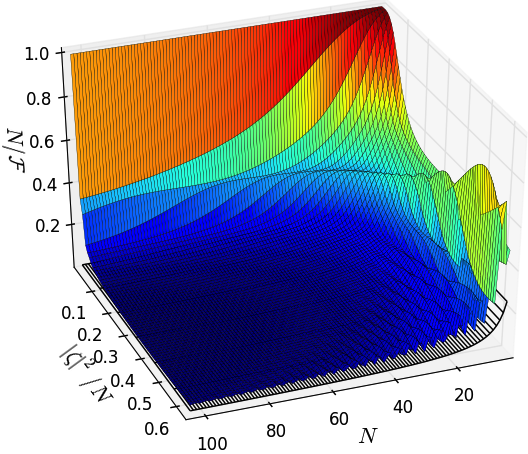}
\caption{(colour online). $N/\mathcal{F} =1$ at $|\zeta|^2 =0$, corresponding to the standard quantum limit. The Heisenberg limit, $N/\mathcal{F} = 1/N$ is marked by the black grid under the coloured surface.}
\label{fig:crescentprecis}
\end{figure}

\begin{figure}[h]
\centering
	\includegraphics[width=75mm]{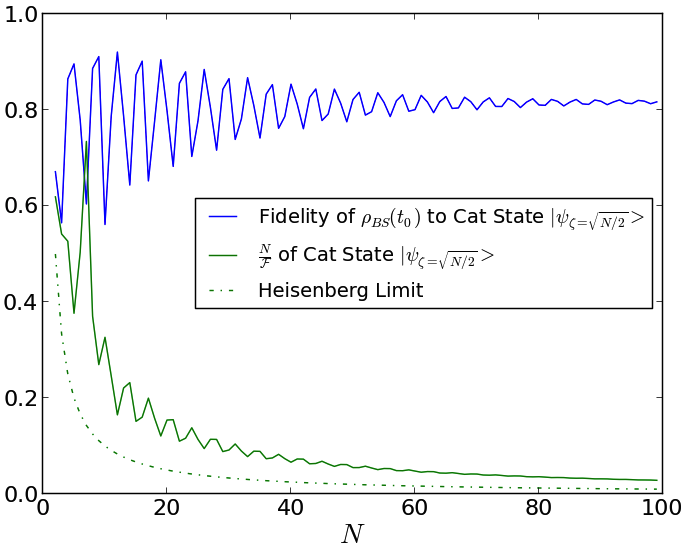} 
	\caption{(colour online). The blue (upper) line is the cross section of figure \ref{fig:3d} at $\abssq{\zeta}/N = 0.5$ and the green (lower) line is the cross section of figure \ref{fig:crescentprecis} at $\abssq{\zeta}/N = 0.5$. Peaks of fidelity coincide with troughs of $N/\mathcal{F}$.}
	\label{fig:xsection}
\end{figure}

In figure \ref{fig:crescentprecis} we plot $N / \mathcal{F}$ against $\abssq{\zeta}/N$ for different values of $N$ up to $N=100$. If $\zeta = 0$, our initial state $\ket{\psi_{\zeta =0} }$ is just a spin coherent state and $ N/F =1$, the standard quantum limit. The Heisenberg limit, $N/\mathcal{F} = 1/N$, is marked in figure \ref{fig:crescentprecis} by a black line for each $N$ (the grid under the coloured contour plot). We see that, especially for large $N$, our cat state $\ket{\psi_{\zeta} }$ can allow for magnetic field sensing close to the Heisenberg limit, even in the $1 \ll\abssq{\zeta}\ll N$ regime in which the cat state emerges from the collapse and revival dynamics. Also in figure \ref{fig:crescentprecis}, we notice the ripples in $N/\mathcal{F}$ at $\abssq{\zeta}/N \approx 0.5$. These ripples are most pronounced for small values of $N$. The green (lower) line in figure \ref{fig:xsection} (which is the cross section of figure \ref{fig:crescentprecis} at $\abssq{\zeta}/N = 0.5$) show that the dips in $N/\mathcal{F}$ coincide with the peaks in fidelity of the big spin to a cat state [the blue (upper) line]. In other words, in this region of parameter space the cat states that are most useful for magnetic field sensing are also the states that can be generated with the highest fidelity by interacting the big spin with the qubit for a time $t_0$ via Hamiltonian (\ref{eq:spinH}). 

\section{Conclusion}
We have considered collapse and revival phenomena for a single spin coupled to a composite big spin, identifying a parameter regime corresponding to conventional qubit-field mode behaviour. Here the evolving system can produce non-classical (cat type) states of the big spin for modest values ($\sim 40$) of $N$, the number of component spins in the big spin. Such states are capable of quantum-enhanced field sensing that approaches the Heisenberg limit. Approximate cat states can also be generated for smaller values of $N$ (e.g. $N=12$) in a different parameter regime. States with such small $N$ values are potentially accessible with current quantum technologies, such as superconducting circuits or multi-spin molecules.

From a practical perspective, these states can be generated from very straightforward initial states: the qubit is initially in a pure state and the big spin is in a separable state of its $N$ component spins with each of the spins aligned (a spin coherent state). Furthermore, any physical realisation for which there is control at the Rabi period time scale should have sufficient time resolution to identify the time(s) at which a cat is generated.

Future work will consider in detail the robustness of the collapse and revival and cat state generation to various forms of decoherence. In addition, in the $\abssq{\zeta}\approx N$ parameter regime it is possible to explore other regimes of ``non-standard'' collapse and revival. These will be discussed in a forthcoming paper \cite{Doo-13}.

\bibliography{refs}

\appendix

\section{}\label{app:A}

The $J_{\pm}$ and $J_z$ operators, restricted to the $j=\frac{N}{2}$ subspace, can be written as

\small\begin{eqnarray} \frac{\hat{J}_{+}}{\sqrt{N}} &=& \sum_{n=0}^{N}\sqrt{( n+1) \left( 1 - \frac{n}{N} \right)} \ket{\frac{N}{2},n + 1 - \frac{N}{2}}\bra{\frac{N}{2},n - \frac{N}{2}} \; ; \nonumber \\  \frac{\hat{J}_{-}}{\sqrt{N}} &=& \sum_{n=0}^{N}\sqrt{n \left( 1 - \frac{n -1}{N} \right)} \ket{\frac{N}{2},n - 1 - \frac{N}{2}}\bra{\frac{N}{2},n - \frac{N}{2}} ;\nonumber \\ J_z &+& \frac{N}{2} = \sum_{n=0}^{N} n \ket{\frac{N}{2},n - \frac{N}{2}}\bra{\frac{N}{2},n - \frac{N}{2}} . \nonumber \end{eqnarray} \normalsize 
The linear map $f$ takes the Dicke state $\ket{\frac{N}{2},n - \frac{N}{2}}$ to the Fock state $\ket{n}$. Taking the $N\to\infty$ limit of $f \frac{J_{\pm}}{\sqrt{N}}f^{\dagger}$ and $f\left( J_z + \frac{N}{2}\right) f^{\dagger}$ we find 

\begin{eqnarray} &\displaystyle\lim_{N\to\infty}& f \frac{J_{+}}{\sqrt{N}}f^{\dagger}  = \sum_{n=0}^{\infty}\sqrt{( n+1)} \ket{n + 1}\bra{n} = \hat{a}^{\dagger} ; \label{eq:HP1} \\  &\displaystyle\lim_{N\to\infty}& f \frac{J_{-}}{\sqrt{N}}f^{\dagger}  = \sum_{n=0}^{\infty}\sqrt{n} \ket{n - 1}\bra{n} = \hat{a} ; \label{eq:HP2} \\ &\displaystyle\lim_{N\to\infty}& f\left( J_z + \frac{N}{2}\right) f^{\dagger} = \sum_{n=0}^{\infty} n \ket{n}\bra{n} = \hat{a}^{\dagger}\hat{a} .  \label{eq:HP3} \end{eqnarray}

The right hand sides of \ref{eq:HP1}, \ref{eq:HP2}, \ref{eq:HP3} of  are exactly the bosonic creation, annihilation, and number operators respectively. (This can also be seen by taking the large $N$ limit of the Holstein-Primakoff transformations \cite{Hol-40}.)

%\begin{equation} \frac{\hat{J}_+}{\sqrt{N}} = \hat{a}^{\dagger}\sqrt{1 - \frac{\hat{a}^{\dagger}\hat{a}}{N}} \quad ; \quad \frac{\hat{J}_-}{\sqrt{N}} =  \sqrt{1 -  \frac{\hat{a}^{\dagger}\hat{a}}{N}} \, \hat{a}. \label{eq:HP}\end{equation}  These transformations relate $\hat{J}_-$ and $\hat{J}_+$ to operators $\hat{a}$ and $\hat{a}^{\dagger}$ that obey the bosonic commutation relation $\left[ \hat{a}, \hat{a}^{\dagger}\right] = 1$ (the operators $\hat{J}_{\pm}$ do not satisfy this relation since $\left[ \hat{J}_-, \hat{J}_{+} \right] = -2\hat{J}_z$).

We now consider the $N\to\infty$ limit of the state $f \ket{N, \frac{\zeta}{\sqrt{N}}}$. We first write $f \ket{N, \frac{\zeta}{\sqrt{N}}}$ as

\begin{eqnarray} f \ket{N, \frac{\zeta}{\sqrt{N}}} &=& \sum_{n=0}^{N} \frac{1}{\left(  1 + \frac{\abssq{\zeta}}{N}   \right)^{N/2}} \sqrt{ {N}\choose{n} }\: \left(\frac{\zeta}{\sqrt{N}}\right)^n \ket{n} \nonumber \\ &=& \sum_{n=0}^{N} \left[ {{N}\choose{n}} (1-p)^{N-n} p^n \right]^{1/2} e^{i \phi n}\ket{n}, \nonumber \end{eqnarray} where $p \equiv \frac{\abssq{\zeta}/N}{1 + \abssq{\zeta}/N}$. The term in the square brackets is the binomial distribution. The \emph{Poisson Limit Theorem} \cite{Pap-02} states that if $N\to\infty$ and $p\to 0$ such that $Np \to \lambda$, then ${{N}\choose{n}} (1-p)^{N-n} p^n \to e^{-\lambda}\frac{\lambda^{n}}{n!}$ in this limit. For our $p = \frac{\abssq{\zeta}/N}{1 + \abssq{\zeta}/N}$, it is clear that when $N\to \infty$ we have $p\to 0$ and $Np \to \abssq{\zeta}$ as required so that

\begin{eqnarray}  \lim_{N\to\infty} f \ket{N, \frac{\zeta}{\sqrt{N}}}  &=& \sum_{n=0}^{\infty} \left[ e^{-\abssq{\zeta}} \frac{|\zeta|^{2n}}{n!} \right]^{1/2} e^{i\phi n} \ket{n} \\ &=& e^{-\abssq{\zeta}/2}\sum_{n=0}^{\infty} \frac{\zeta^n}{\sqrt{n!}}\ket{n} \\ &=& \ket{\zeta} , \end{eqnarray} the coherent state of the field mode.

%Since our Hamiltonian in the interaction picture (on resonance) is $H_I = \frac{\lambda}{\sqrt{N}} \left(  J_+ \sigma_- + J_- \sigma_+  \right)$, the unitary time evolution operator is \begin{eqnarray}  U(t) = e^{-itH_I} = \sum_{k=0}^{\infty} \frac{ \left( -it\lambda\right)^k }{k!}   \left(  J_+ \sigma_- + J_- \sigma_+  \right)^k .  \end{eqnarray} Using the fact that $\sigma_- \sigma_- = \sigma_+ \sigma_+ = 0$, we can write this as \begin{widetext} \begin{eqnarray} U(t) &=& \sum_{k=0}^{\infty} \frac{ \left( -it\lambda\right)^k }{k!}   \left(  \underbrace{J_+ J_- J_+...}_{k \mbox{ times}} \otimes \underbrace{\sigma_- \sigma_+ \sigma_-...}_{k \mbox{ times}} \: + \: \underbrace{J_- J_+ J_-...}_{k \mbox{ times}} \otimes \underbrace{\sigma_+ \sigma_- \sigma_+...}_{k \mbox{ times}}  \right)  \\ &=& ...  \\ &=& \cos \left( \lambda t \sqrt{\frac{J_+ J_-}{N}} \right) \ket{1}\bra{1} + \cos \left( \lambda t \sqrt{\frac{J_- J_+}{N}} \right) \ket{0}\bra{0} -  \nonumber  \\  & & \quad  - i \sin \left( \lambda t \sqrt{\frac{J_+ J_-}{N}} \right) \left( J_+ J_- \right)^{-1/2} J_+  \ket{1}\bra{0} - i \sin \left( \lambda t \sqrt{\frac{J_- J_+}{N}} \right) \left( J_- J_+ \right)^{-1/2} J_- \ket{0}\bra{1}  \end{eqnarray} \end{widetext} Consider this evolution operator acting on the initial state $\ket{N,\zeta}\ket{D_{\pm}(0)}$

\end{document}